\documentclass[iop, twocolumn]{aastex6}
\usepackage{amsmath}     
\usepackage{wasysym}           
\usepackage{graphicx}
\usepackage{amssymb}
\usepackage{epstopdf}
\usepackage{mathrsfs}
\usepackage{natbib}
\usepackage{color}
\usepackage{lipsum}

\begin{document}
\title{Evidence for universality in the initial planetesimal mass function}
\author{Jacob B. Simon\altaffilmark{1,2,3}, Philip J. Armitage\altaffilmark{1,3,4}, Andrew N. Youdin\altaffilmark{5} and Rixin Li\altaffilmark{5}}

\altaffiltext{1}{JILA, University of Colorado and NIST, 440 UCB, Boulder, CO 80309-0440}
\altaffiltext{2}{Department of Space Studies, Southwest Research Institute, Boulder, CO 80302}
\altaffiltext{3}{Kavli Institute for Theoretical Physics, UC Santa Barbara, Santa Barbara, CA 93106}
\altaffiltext{4}{Department of Astrophysical and Planetary Sciences, University of Colorado, Boulder}
\altaffiltext{5}{Department of Astronomy and Steward Observatory, University of Arizona, 933 North Cherry Avenue, Tucson, AZ 85721}

\email{jbsimon.astro@gmail.com}

\begin{abstract}
Planetesimals may form from the gravitational collapse of dense particle clumps initiated by 
the streaming instability. We use simulations of aerodynamically coupled gas-particle mixtures to 
investigate whether the properties of planetesimals formed in this way depend upon the 
sizes of the particles that participate in the instability. Based on three high resolution simulations that span 
a range of dimensionless stopping time $6 \times 10^{-3} \leq \tau \leq 2$ no statistically 
significant differences  in the initial planetesimal mass function are found. The mass functions 
are fit by a power-law, ${\rm d}N / {\rm d}M_p \propto M_p^{-p}$, with $p=1.5-1.7$ and errors 
of $\Delta p \approx 0.1$. Comparing the particle density fields prior to collapse, we find 
that the high wavenumber power spectra are similarly indistinguishable, though the large-scale
geometry of structures induced via the streaming instability is significantly different between
all three cases. We interpret the results as evidence for a near-universal slope to the mass function, arising from the small-scale structure of streaming-induced turbulence. 
\end{abstract}


\section{Introduction}
The streaming instability \citep{youdin05} leads to clustering of aerodynamically coupled solids across a 
broad range of protoplanetary disk conditions \citep{johansen07b,bai10}. 
Because the physical origin of the instability is closely tied to that of radial drift \citep{whipple72,weidenschilling77}---the 
most widespread barrier to growth beyond small macroscopic sizes---there is a compelling circumstantial 
argument that streaming plays a major role in planetesimal formation. The simplest scenario is that 
coagulation and radial drift lead to local conditions that trigger the streaming instability, which then 
forms dense particle clumps that collapse gravitationally to form planetesimals \citep{johansen07}. 
More involved variants, in which persistent or transient disk structures (ice lines, zonal flows, vortices, 
dead zone edges, etc.) are pre-requisites for streaming-initiated collapse, are 
also possible \citep{johansen14,armitage15}.

The parameters that determine the operation of the streaming instability include the particle size 
(measured via the dimensionless stopping time $\tau$), the ratio of the solid to gas surface density 
$Z$, which we colloquially refer to as ``metallicity", and the degree of pressure support in the gas. These parameters vary 
with radius in the disk, and a successful theory of planetesimal 
formation must therefore work across a range of starting conditions. This requirement 
is readily satisfied by the streaming instability at the linear level, where a broad 
array of gas / particle mixtures are linearly unstable \citep{youdin05}. At the non-linear level, 
simulations for $\tau \sim 1$ show that the metallicity $Z$ needs to exceed the nominal dust to gas ratio of $0.01$ before the instability produces the strong clumping that precedes collapse \citep{johansen09}, but 
provided high $Z$ can be reached particles with $10^{-3} \lesssim \tau \lesssim 5$ are viable 
progenitors \citep{carrera15,yang16}.\footnote{The precise values of $Z$ and $\tau$ for which the streaming instability
operates also depend on the radial pressure gradient \citep{bai10b}.} The allowable range of $\tau$ may be more strongly restricted 
at the low end by intrinsic turbulence in the gas, though recent theoretical results \citep{simon15} and 
observations \citep{flaherty15} suggest that at least some disks may be less turbulent than 
was previously thought.

In this {\em Letter} we investigate whether the outcome of streaming-initiated collapse is 
universal, in the sense of forming an initial mass function of planetesimals whose shape 
is independent of the aerodynamic properties of the particles that participate in the instability. 
High resolution simulations have shown that the prediction for the initial mass 
function is a power-law, with a cut-off at high masses that is set by the local mass of 
solids that participates in the instability \citep{johansen12,simon16,schafer17}. The 
existing simulations, however, have focused on a range of $\tau$ that is much 
smaller than that realized in actual disks. Here, we present results from simulations 
that span a range of $\tau$ between 0.006 and 2. We analyze the non-linear particle structures and 
the mass function of collapsed objects produced in the simulations, and show that any 
deviations from universality across this range of parameters must be small.

\section{Methods}
Our results are based on supplementing the highest resolution simulation ($512^3$ gas zones, 
$1.536 \times 10^8$ particles) from \cite{simon16} with two further calculations initialized 
with differing values of the stopping time and metallicity. We work in a locally Cartesian 
``shearing box" with radial, azimuthal and vertical co-ordinates $(x,y,z)$, and model 
the fluid as an isothermal gas with pressure $P=\rho c_s^2$. In the rotating frame 
the hydrodynamics of the gas is described by,
\begin{eqnarray}
 \frac{\partial \rho}{\partial t} + \nabla \cdot \left( \rho {\bf u} \right) & = & 0, \\
 \frac{\partial \rho {\bf u}}{\partial t} + \nabla \cdot \left( \rho {\bf u} {\bf u} + P {\bf I} \right) & = &  
 3 \rho \Omega^2 x \hat{\bf x} - \rho \Omega^2 z \hat{\bf z} \nonumber \\
 & - & 2 {\bf \Omega} \times \rho {\bf u}
 + \rho_p \frac{ {\bf v} - {\bf u} }{t_{\rm stop}}. 
\end{eqnarray} 
Here ${\bf u}$ and ${\bf v}$ are velocities of the gas and particle fluids, respectively, ${\bf I}$ is the identity matrix, 
$\rho_p$ is the particle density, $t_{\rm stop}$ is the dimensional stopping time, and $\Omega$ 
the angular velocity which is assumed to be Keplerian.

The solids are represented as discrete super-particles \citep{youdin07a} $i = 1, ..., N$. In the 
shearing box frame they are subject to the fictitious forces arising from the rotating co-ordinate 
system, the vertical component of stellar gravity, a force representing the dynamical effect of the 
radial pressure gradient, and self-gravity,
\begin{eqnarray}
 \frac{{\rm d} {\bf v}_i}{{\rm d} t} = 2 {\bf v}_i \times {\bf \Omega} + 3 \Omega^2 x \hat{\bf x} 
 - \Omega^2 z \hat{\bf z} \nonumber \\
 -\frac{{\bf v}_i - {\bf u}}{t_{\rm stop}} - {\bf F}_{\rm p} + {\bf F}_{\rm g}.
\end{eqnarray} 
A shearing box representation necessarily has no radial pressure gradient across the 
domain. To induce radial drift we instead impose a constant inward force on the particles 
${\bf F}_{\rm p} = -2 \eta v_K \Omega \hat{\bf x}$, where $\eta v_K$ is the deviation 
from Keplerian orbital velocity due to radial pressure gradients in the physical system.
The term ${\bf F}_{\rm g}$ is the particle self-gravity term, derived from a solution to 
Poisson's equation,
\begin{eqnarray}
 \nabla^2 \Phi_{\rm p} & = & 4 \pi G \rho_{\rm p} \\
 {\bf F}_{\rm g} & = & - \nabla \Phi_{\rm p}.
\end{eqnarray} 
Where necessary, interpolation is used to map fluid quantities defined on a fixed grid 
to the locations of individual particles, and vice versa \citep{bai10,simon16}.

The coupled particle-gas system is solved using the {\sc athena} code \citep{stone08}, 
in practice in a slightly different form that subtracts off the local orbital advection 
velocity. Established methods are used to implement the shearing boundary 
conditions \citep{stone10}, particle integration \citep{bai10} and particle 
self-gravity \citep{koyama09,simon16}, which is calculated using a Particle-Mesh scheme.

The streaming instability is characterized by the dimensionless stopping time of the 
participating particles,
\begin{equation}
 \tau \equiv t_{\rm stop} \Omega,
\end{equation}
the local metallicity, defined via the ratio of particle to gas surface densities,
\begin{equation}
 Z \equiv \frac{\Sigma_p}{\Sigma_g},
\end{equation}
a pressure gradient parameter,
\begin{equation}
 \Pi \equiv \eta v_K / c_s,
\end{equation}
and a parameter describing the the relative strength of self-gravity versus tidal shear,
\begin{equation}
 \tilde{G} \equiv \frac{4 \pi G \rho_0}{\Omega^2}.
\end{equation}
Here, $\rho_0$ is the mid-plane gas density. We fix $\Pi = 0.05$ and $\tilde{G} = 0.05$, equivalent
to a gaseous Toomre $Q \approx 32$. Our three runs sample a range of stopping times and metallicities,
\begin{itemize}
\item
$\tau = 0.3$, $Z=0.02$. This is the highest resolution run from \cite{simon16}.
\item
$\tau = 2$, $Z=0.1$. This 
is a distinct parameter set that remains relatively easy to simulate. Radial drift 
means that it is hard to attain the $\tau > 1$ regime in real disks \citep{birnstiel12}, but it could be 
physically relevant if solids grow in particle traps in the outer disk \citep{pinilla12}.
\item
$\tau = 0.006$, $Z=0.1$. These parameters approach those expected for 
mm-cm sized solids that drift and pile-up in the dense region interior to 
the snow line \citep{youdin04}.
\end{itemize}
The results of \cite{carrera15} and \cite{yang16} imply that all three runs ought to result in strong 
clustering that is unstable to gravitational collapse, and this expectation is confirmed.

All other numerical details follow those described in \cite{simon16}. The simulations use 
a cubical box of size $(0.2 H)^3$, where $H = c_s / \Omega$, $512^3$ grid zones for the 
gas, and $N = 1.536 \times 10^8$ particles.

\section{Mass function of collapsed structures}
In common with most prior work, the simulations are run in two steps. Initially, we evolve the 
aerodynamically coupled system in the absence of self-gravity. When the 
system has attained a saturated state (defined as when the maximum particle mass density is statistically
constant in time, with moderate stochastic fluctuations about this constant value) self-gravity is turned on and collapse to planetesimals 
proceeds. For the $\tau=0.006, 0.03,$ and 2 runs, self-gravity is switched on at $346.3 \ \Omega^{-1}$, $110 \ \Omega^{-1}$, and
$37 \ \Omega^{-1}$, (at which point the RMS scale height for the particles are $0.01H, 0.006H$, and $0.002H$) respectively. 
This two step procedure is followed to reduce the computational expense of the runs. It is 
justified by tests (admittedly at lower resolution) that find no evidence that the outcome depends 
on the timing of self-gravity onset \citep{simon16}.

In addition,  we have chosen to stop each simulation when $\sim 100$ planetesimals have formed. This number is a compromise between statistical precision and computational cost. 
The chosen stopping times are $468.2 \ \Omega^{-1}$ $117.6 \ \Omega^{-1}$, and $45.8\  \Omega^{-1}$ for the 
$\tau=0.006, 0.03,$ and 2 runs, respectively. Upon examination of the particle mass
surface density, it is possible that, at least for the $\tau = 0.006$ and 0.3 runs, further planetesimal
formation will occur.  However, based on the evolution of the power law index and minimum and maximum planetesimal
masses (as discussed further below), we
believe that further planetesimal formation will not appreciably change the mass functions.

\begin{figure*}
\includegraphics[width=\textwidth]{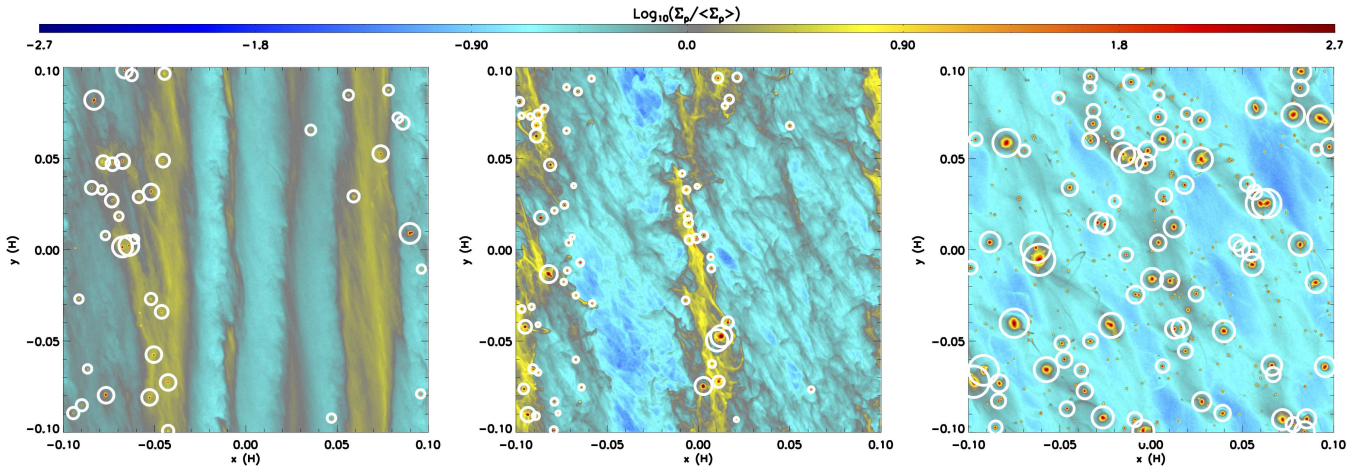}
\caption{The surface density of solids in the $x$-$y$ (orbital) plane, shown after self-gravity has been 
turned on and dense clumps have collapsed. From left to right, the panels depict runs with 
$\tau=0.006$, $\tau=0.3$ and $\tau=2$, at times $105.4\Omega^{-1}$, $7.6\Omega^{-1}$ and $8.8\Omega^{-1}$  after self-gravity has been switched on, respectively. 
White circles depicts the Hill spheres for a subset of the identified planetesimals. Both the solid surface density structure and the initial 
positions of planetesimals become more axisymmetric at low values of the stopping time.}
\label{fig_snapshots}
\end{figure*}

Figure~\ref{fig_snapshots} shows projections of the surface density of solid material for
the three simulation runs in the orbital plane after dense clumps have formed. Visually, they are quite distinct. 
The prominence of axisymmetric bands in the solid surface density decreases with 
increasing $\tau$, and notably less material collapses promptly into bound clumps 
for the $\tau=0.006$ run.  In the smaller $\tau$ runs, the planetesimals form primarily in
two azimuthally extended bands.  In contrast, the planetesimals in the $\tau = 2$ run fill the box much
more uniformly.  

\begin{deluxetable}{llll}
\tablecaption{Summary of simulation results\label{table1}}
\tablehead{\colhead{$\tau$} & \colhead{$Z$} & \colhead{$p$\tablenotemark{a}} & \colhead{$m$\tablenotemark{b}}}
\startdata
0.006 & 0.1 & 1.73 $\pm$ 0.11 & 1.50 $\pm$ 0.02 \\
0.3 & 0.02 &  1.61 $\pm$ 0.08 & 1.65 $\pm$ 0.10 \\
2.0 & 0.1 & 1.54 $\pm$ 0.04 & 1.62 $\pm$ 0.11 \\
\enddata
\tablenotetext{a}{Slope of the differential mass function}
\tablenotetext{b}{Slope of the power spectrum of solids within the range $100 < k H/2\pi < 1250$ 
prior to turning on self-gravity}
\end{deluxetable}

\begin{figure*}
\includegraphics[width=0.5\textwidth]{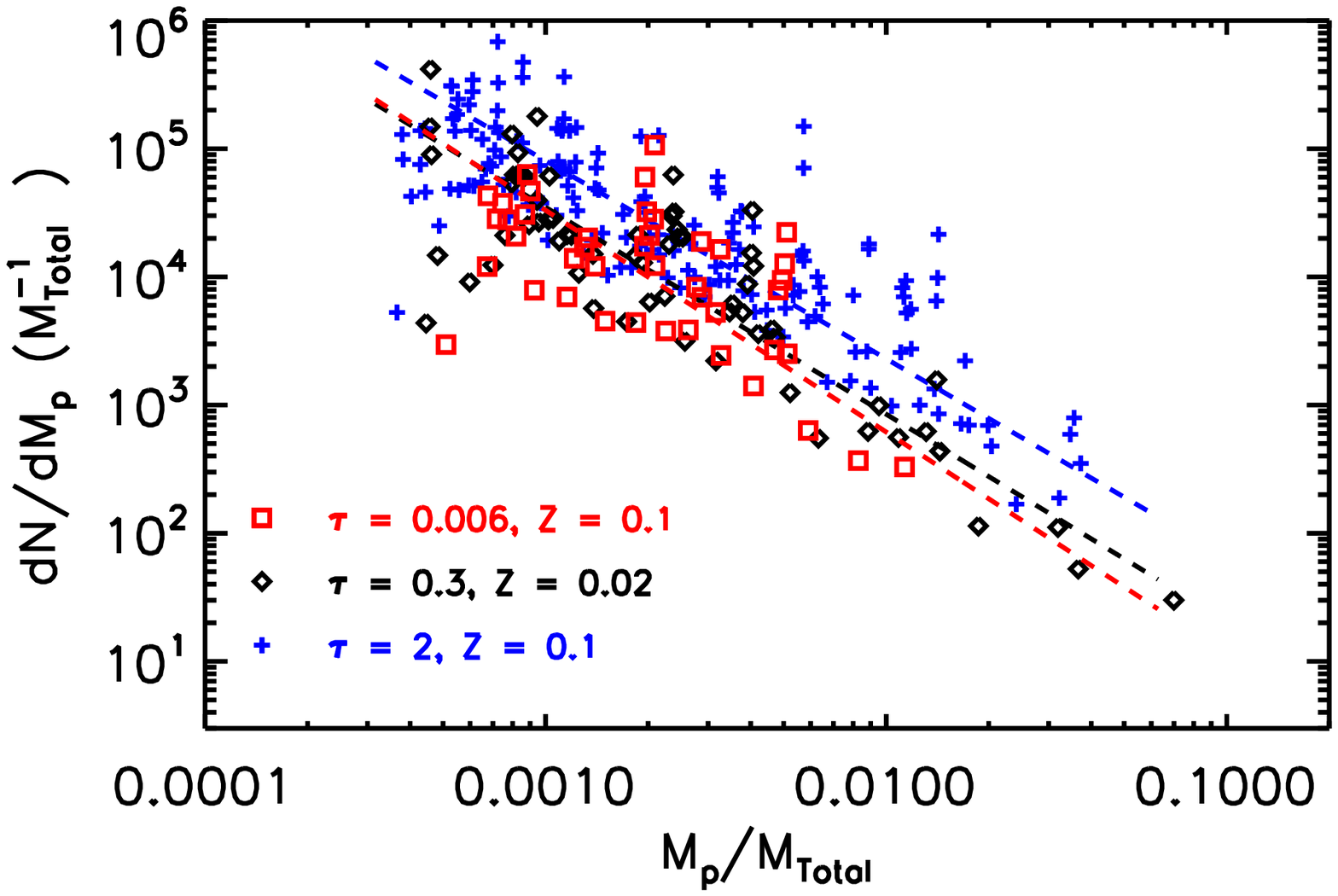}
\includegraphics[width=0.5\textwidth]{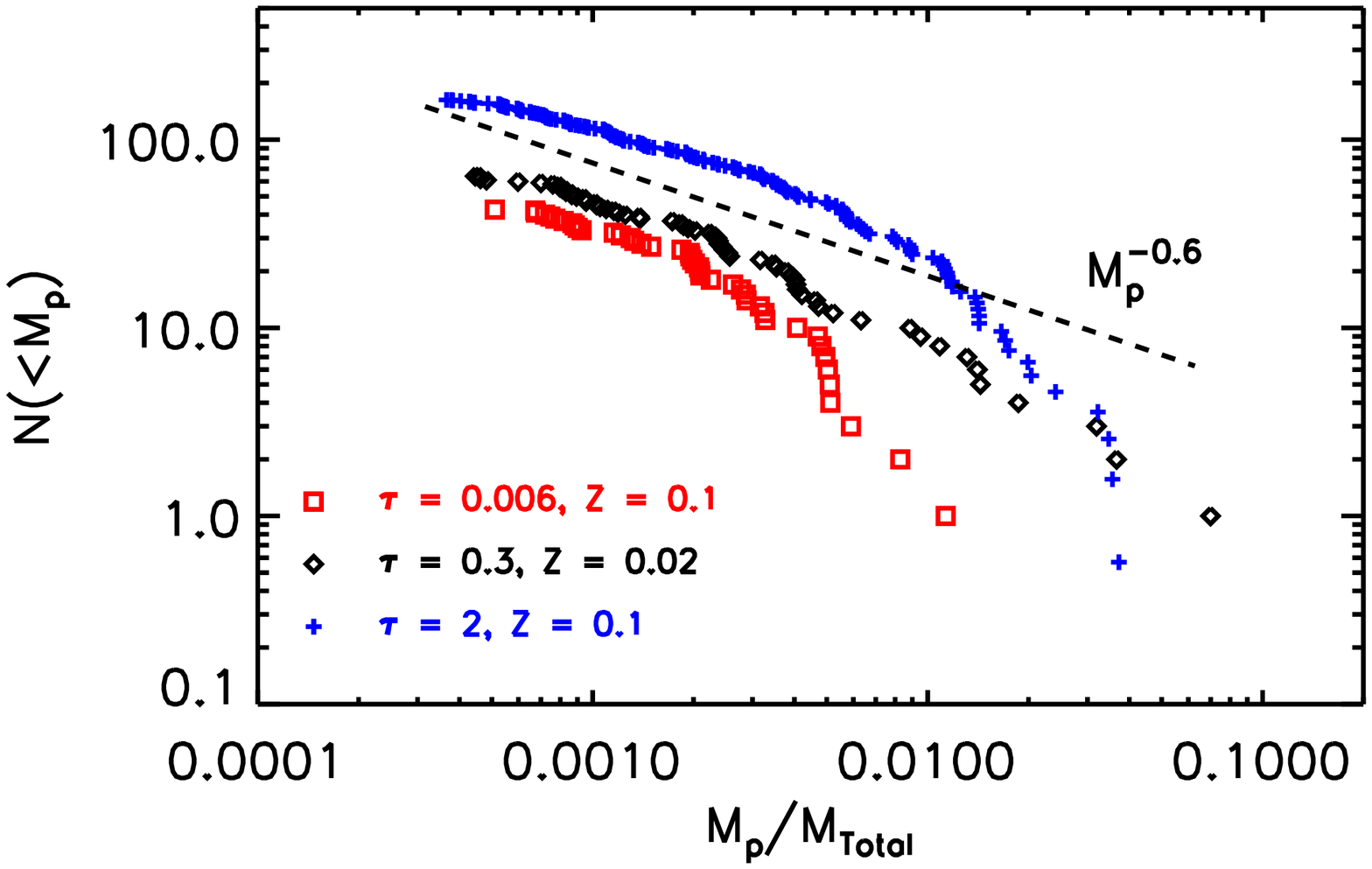}
\caption{The differential (left) and cumulative (right) initial planetesimal mass function derived from the simulations. In the left plot, the points show 
unsmoothed ``local" estimates of the mass function, and the lines show best-fit solutions 
derived using a maximum likelihood estimator assuming a power-law form. The simulation with the 
smallest particles ($\tau = 0.006$, red) forms a significantly smaller total mass of planetesimals 
throughout the duration of the run, but no significant differences in the slope of the derived mass 
function are observed. The agreement between the points on the low mass end of the cumulative function
and the $M_p^{-0.6}$ power law (dashed line) demonstrates that at small masses, the
mass function is well represented by the single power law fit that we have calculated.}
\label{fig_mass_function}
\end{figure*}

From these snapshots, we use a clump-finding routine to identify and measure the 
bound masses $M_{{\rm p},i}$ of collapsed objects \citep[we use an identical 
algorithm to][]{simon16}. For visual purposes, we compute an unsmoothed 
estimate of the differential mass function at masses corresponding to each 
planetesimal,
\begin{equation}
 \left. \frac{{\rm d}N}{{\rm d} M_{\rm p}} \right|_i = \frac{2}{M_{{\rm p}, i+1} - M_{{\rm p}, i-1}},
\end{equation} 
The resulting mass functions are plotted in the left plot of Figure~\ref{fig_mass_function}. The 
shape of the mass functions is generally consistent with a single power-law for 
all three runs, though, as we discuss in more detail
below, there is likely a cut-off at high mass.  The single power law fit
is clearest in the intermediate and high-$\tau$ runs 
which form planetesimals with a broader range of masses.
Assuming that the data is indeed drawn from a power-law 
distribution, ${\rm d}N / {\rm d}M_p \propto M_{\rm p}^{-p}$, we proceed to 
estimate the index $p$ and error $\sigma$ using a maximum likelihood 
estimator \citep{clauset09}. Given $n$ planetesimals with masses 
$M_{{\rm p},i} \geq M_{\rm p, min}$, we have,
\begin{eqnarray}
 p & = & 1 + n \left[ \sum_{i=1}^{n} \ln \left( \frac{M_{{\rm p},i}}{M_{\rm p,min}} \right) \right]^{-1}, \\
 \sigma & = & \frac{p-1}{\sqrt{n}}.
\end{eqnarray} 
The derived slopes and their associated errors are given in Table~\ref{table1}. 
Within 1-1.5~$\sigma$ all three runs are consistent with $p=1.6$, as 
found both in our previous work \citep{simon16} and in \cite{johansen15}.  It is also worth noting
that despite the consistent value of $p$, the total fraction of mass converted to planetesimals varies strongly with stopping time; $\sim 10\%$, $40\%$, and $70\%$
of the total mass in solids is converted to planetesimals for the $\tau = 0.006, 0.3,$ and 2 simulations, respectively.  For the smaller mm--cm sized solids
present in the inner regions of disks, our results suggest a relatively low planetesimal formation efficiency in these regions.

We also calculate the cumulative mass function, as shown in the right plot of Fig.~\ref{fig_mass_function}. 
A power law index of $-0.6$ clearly agrees with the mass function slope for low mass objects. This suggests
that the differential mass functions are well fit by a single power law because there is a significantly larger number of planetesimals
at small masses, thus weighting the fit towards the small mass end.
Therefore, while in reality the mass distribution will have a cut off at some high
mass value (which depends on $\tau$), the differential distribution can be well fit
by a single power law that is representative of masses away from the cut off.

\begin{figure*}
\includegraphics[width=0.34\textwidth]{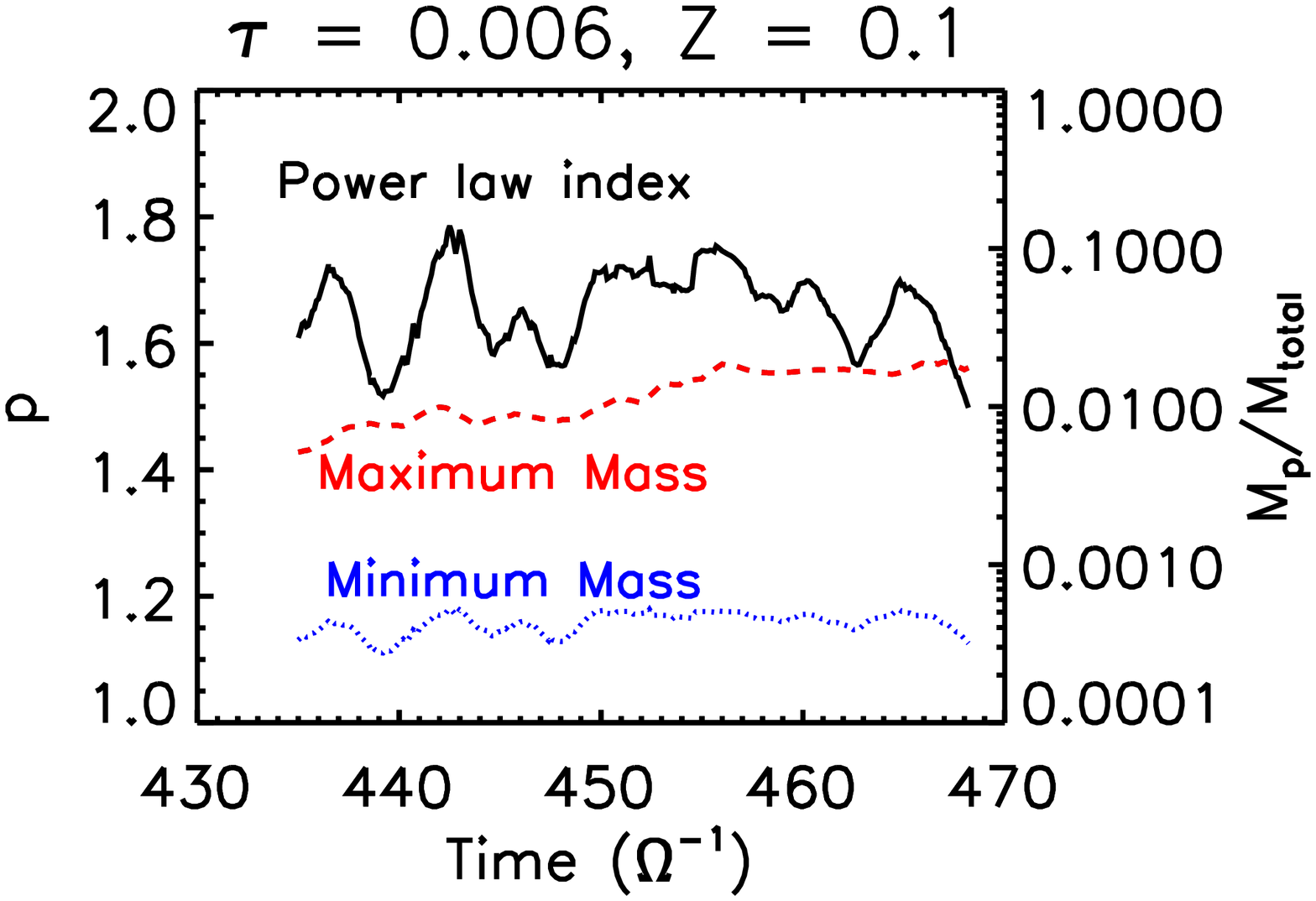}
\includegraphics[width=0.34\textwidth]{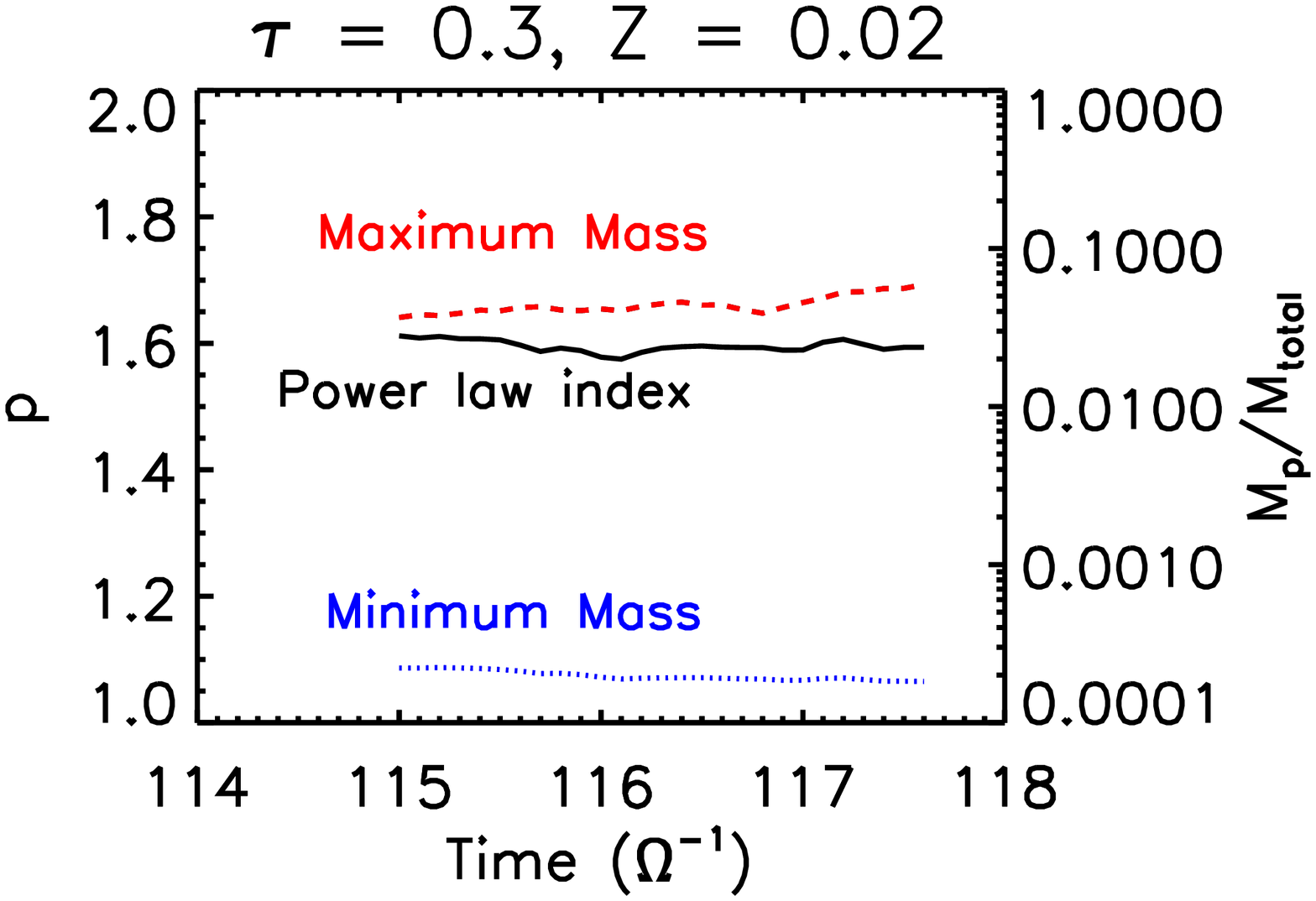}
\includegraphics[width=0.34\textwidth]{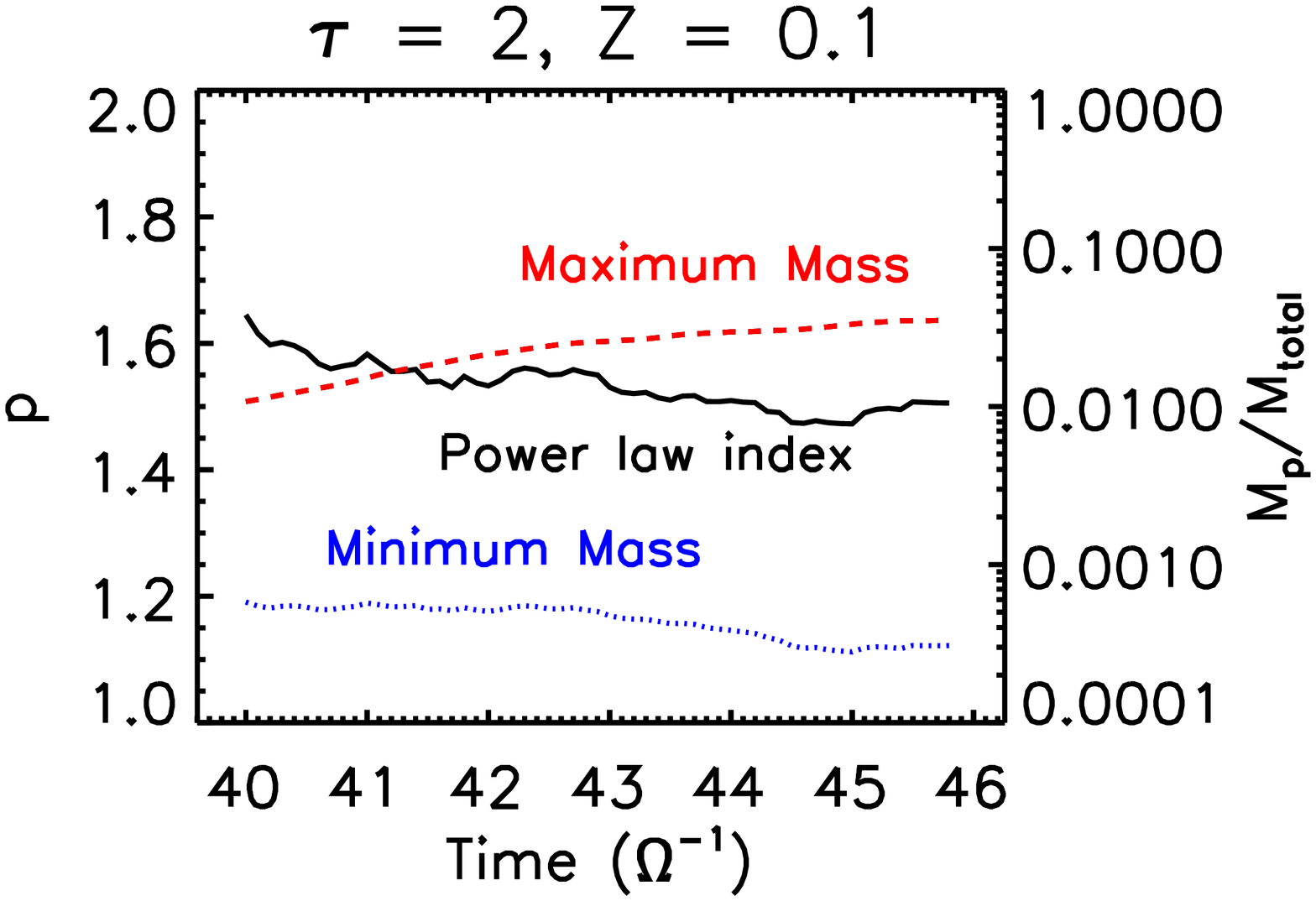}
\caption{Time evolution of the power law index (solid, black line), maximum planetesimal mass (dashed, red line), and minimum planetesimal mass
(dotted, blue line) for the $\tau = 0.006$ (left), $\tau = 0.3$ (middle), and $\tau = 2$ (right) simulations for a period of
time after planetesimals have formed.   Both the power law index, $p$, and
the minimum planetesimal mass are relatively constant in time, whereas the maximum planetesimal mass grows slowly, presumably
due to continued accretion of small particles and/or mergers with smaller planetesimals.}
\label{mft}
\end{figure*}

Finally, we have also examined the evolution of $p$ and the maximum and
minimum planetesimal masses (in all the three simulations) for times after which
a significant number of planetesimals have formed; the results are shown in Fig.~\ref{mft}. Note that due to 
limitations with the clump finder algorithm (as described more fully in \citealt{simon16}),
we smoothed the curves (with a box car average) to remove noise associated
with this algorithm.  As the figure shows, $p$ and the minimum planetesimal mass are relatively constant in time
\footnote{We do observe a slight decrease in the minimum mass in some instances, a behavior that
could arise from a combination of processes, including tidal stripping of smaller bodies by large planetesimals,
fragmentation of smaller bodies, and/or the preferential production of smaller planetesimals as the reservoir of 
particles from which to produce planetesimals decreases in size \citep{johansen15}.  Elucidating the nature
of this behavior requires a more sophisticated clump-finding algorithm and will thus be reserved for future work.},
whereas the maximum planetesimal mass grows by a factor of order unity in each simulation, likely due to accretion
of smaller particles and/or mergers with smaller planetesimals. Despite this
growth the mass function does not evolve appreciably once planetesimals have formed.

Our ability to directly measure any potential 
dependence of the mass function on particle properties is limited by the 
relatively small number of collapsed clumps, but across the range of $\tau$ 
considered, we can bound deviations at the level of $\Delta p \lesssim 0.1-0.2$. 
We can therefore exclude, with moderately high confidence, the possibility 
that the streaming instability might result in a steep mass function ($p > 2$) 
with most of the mass in the smallest planetesimals. 

\section{Particle clustering prior to collapse}
The similarity in the mass functions from the different runs is somewhat surprising 
given the visual differences in the large-scale particle structures that are 
collapsing (Figure~\ref{fig_snapshots}). To identify possible differences 
on smaller spatial scales, we compute the power spectra of the pre-collapse 
particle density fields. The power spectrum maps uniquely to the mass function  
in the case where the density is a gaussian random field \citep{press74}. In the more 
complex case relevant here we use the power spectra only to test whether the 
non-linear structures produced by the streaming instability {\em prior} to the onset 
of self-gravity are indifferent to the particle size.

\begin{figure}
\includegraphics[width=0.5\textwidth]{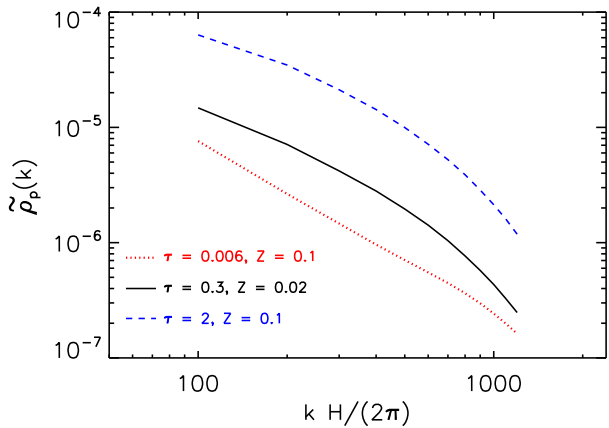}
\caption{The power spectrum of the particle density field, computed via interpolation on to the 
hydrodynamic grid in a fixed (across runs) slice centered on the disk mid-plane. The power 
spectra are computed {\em prior} to turning on self-gravity, in the saturated state of the streaming
instability. At high $k$ the slopes for the $\tau=0.3$ (black, solid) and $\tau=2$ (blue, dashed) runs are 
the same, while the run with smaller particles ($\tau=0.006$; red, dotted) is significantly flatter.}
\label{fig_power_spectrum}
\end{figure}

Figure~\ref{fig_power_spectrum} shows the three dimensional power spectra $\tilde{\rho}_p(k)$ 
computed from a time slice just before self-gravity is switched on. The thickness of 
the particle layer at this stage varies substantially with $\tau$ (higher $\tau$ allows 
for greater settling). To minimize artifacts in the power spectra created by the different 
thicknesses, we compute $\tilde{\rho}_p(k)$ from the interpolated density field $\rho ( {\bf x} )$ 
in a mid-plane slice whose thickness is chosen to be smaller than the scale height 
of the {\em thinnest} particle layer (for $\tau=2$).  The three-dimensional power spectra
are then averaged over shells of constant $|{\bf k}|$. 

Up to normalization differences the power spectra for all three runs display 
a high level of similarity. From calculating the largest Hill radius in each run, 
we expect scales of $k H/2\pi \gtrsim 100$ to seed the collapse.\footnote{$k H/2\pi = 100$
is also the minimum $k$ allowed by choosing a thin $z$ layer}.  Fitting a power-law, $\tilde{\rho}_p(k) \propto k^{-m}$, 
to the data at $k H/2\pi > 100$, we find that $m \approx 1.5$--$1.7$ with reasonably large
errors for the larger two $\tau$ values,\footnote{This error partially results from fitting
a single slope to a spectra that deviates from a simple power law.} 
with the precise values and their associated errors shown in Table~\ref{table1}. 
All three runs yield statistically consistent slopes, suggesting that the 
near-identical mass functions formed from those runs result from commonality in the 
small-scale particle structures formed in the non-self-gravitating non-linear evolution of the 
streaming instability.   It should be noted, however, that from a visual inspection (Fig.~\ref{fig_power_spectrum}) 
the shape and slope of the power spectrum for $\tau = 0.006$ is significantly different (and better fit
with a single power law) than for the higher $\tau$ cases, even though their fitted power-law slopes
are statistically equal.  It is possible that any such differences in the power spectra translate 
into differences in the mass function at a level that is smaller than our current measurement 
uncertainties. 

\section{Discussion}
We have presented results from a small number of high resolution simulations of the 
streaming instability \citep{youdin05} that model the gravitational collapse of over-dense 
structures toward planetesimals. The simulations were initialized with values of the 
dimensionless stopping time and local metallicity that promote strong clustering and prompt collapse \citep{carrera15,yang16}. For 
$0.006 \leq \tau \leq 2$ we find that the power law part of the resulting mass 
function (${\rm d}N / {\rm d}M_p \propto M_p^{-p}$)  has an approximately universal 
slope, $p \simeq 1.6$, consistent with that measured previously for particle sizes in 
the middle of this range \citep{johansen12,simon16}.

The similar planetesimal mass distributions in the different runs is somewhat surprising, given differences in the larger scale geometry 
of the particle clustering. However, this similarity is consistent with the approximately equal slopes in the power spectra of particle clustering on the smaller scales relevant to collapse. It is possible that the observed differences in the shape of the power spectra would translate into small deviations from universality, but within the uncertainties outlined here, our results strongly support a top-heavy mass function for planetesimals.  This finding is consistent with previous streaming instability simulations, but is now shown over a broader range of $\tau$--$Z$ parameter space.

Our ability to directly determine the 
predicted initial mass function is limited by small number statistics. The statistics 
could be improved by co-adding samples derived from independent runs, by 
increasing the spatial resolution, or by increasing the domain size. 
In \cite{simon16}, we showed that the planetesimal mass distribution is essentially independent of the time at which self-gravity is turned on in the simulation.  However this independence still remains to be tested at higher resolutions and across the broader ranges of parameters considered here.

Where gravitational collapse is encountered in other astrophysical settings, notably in 
the hydrodynamic formation of stars \citep{bastian10} and in the collision-less collapse 
of dark matter haloes \citep{navarro97}, it is known to exhibit universal features.
Planetesimals may in principle form from gravitational collapse 
via other routes, for example from the direct collapse of dense particle 
layers \citep{goldreich73,youdin02,shi13}, secular gravitational instability \citep{youdin11,takahashi14} or when vortices accumulate solids  
\citep{barge95,raettig15}. It is of interest to explore whether these processes 
lead to similar or identical top-heavy power-law mass functions to those found here, 
and hence whether constraints on planetesimal formation from the asteroid \citep{morbidelli09} 
and Kuiper belts \citep{nesvorny10} test specifically the streaming instability or rather a broader class 
of gravitational collapse scenarios.  On the other hand, if the mass function is indeed
intimately coupled to the non-linear state of the streaming instability, turbulence driven by other means
\citep[e.g., the magnetorotational instability,][]{balbus98} and imposed onto the streaming instability
may fundamentally alter the mass function of planetesimals.

\acknowledgments
JBS and PJA acknowledge support from NASA through 
grants NNX13AI58G and NNX16AB42G, and from the NSF 
through grant AST 1313021. ANY acknowledges support from the NSF through grant AST~1616929. 
We thank Doug Lin and Katherine Kretke for useful discussions regarding this work. 
We also thank the anonymous referee whose comments greatly improved the quality of this paper.
The computations were performed on Stampede and Maverick at the Texas Advanced Computing Center through XSEDE grant TG-AST120062.
We thank the Kavli Institute for Theoretical Physics, 
supported in part by the National Science Foundation under Grant No. NSF PHY-1125915, 
for hospitality during the completion of the paper.

\end{document}